\documentclass[aps,groupedaddress,superscriptaddress,amsmath,amssymb,twocolumn,prb]{revtex4-2}
\usepackage{amsmath,amssymb,amsfonts}
\usepackage{algorithmic}
\usepackage{graphicx}
\usepackage[normalem]{ulem}
\usepackage{xcolor}
\usepackage{listings}
\usepackage{hyperref}

\begin{document}

\title{Parallel measurements of vibrational modes in a few-layer graphene nanomechanical resonator using software-defined radio dongles}
\author{Heng Lu}
\affiliation{School of Optoelectronic Science and Engineering \& Collaborative Innovation Center of Suzhou Nano Science and Technology, Soochow University, Suzhou, 215006, China}
\author{Chen Yang}
\affiliation{School of Optoelectronic Science and Engineering \& Collaborative Innovation Center of Suzhou Nano Science and Technology, Soochow University, Suzhou, 215006, China}
\author{Ye Tian}
\affiliation{School of Optoelectronic Science and Engineering \& Collaborative Innovation Center of Suzhou Nano Science and Technology, Soochow University, Suzhou, 215006, China}
\author{Jue Wang}
\affiliation{Department of Child Care, Suzhou Municipal Hospital, Suzhou, 215006, China}
\author{Ce Zhang}
\affiliation{School of Optoelectronic Science and Engineering \& Collaborative Innovation Center of Suzhou Nano Science and Technology, Soochow University, Suzhou, 215006, China}
\author{Yubin Zhang}
\affiliation{School of Optoelectronic Science and Engineering \& Collaborative Innovation Center of Suzhou Nano Science and Technology, Soochow University, Suzhou, 215006, China}
\author{FengNan Chen}
\affiliation{School of Optoelectronic Science and Engineering \& Collaborative Innovation Center of Suzhou Nano Science and Technology, Soochow University, Suzhou, 215006, China}
\author{Ying Yan}
\affiliation{School of Optoelectronic Science and Engineering \& Collaborative Innovation Center of Suzhou Nano Science and Technology, Soochow University, Suzhou, 215006, China}
\affiliation{Engineering Research Center of Digital Imaging and Display, Ministry of Education, Soochow University, Suzhou, 215006, China}
\author{Joel Moser}
\email{j.moser@suda.edu.cn}
\affiliation{School of Optoelectronic Science and Engineering \& Collaborative Innovation Center of Suzhou Nano Science and Technology, Soochow University, Suzhou, 215006, China}

\begin{abstract}
Software-defined radio dongles are small and inexpensive receivers well known to amateur radio enthusiasts. When connected to an antenna, they enable monitoring of a wide range of the radio spectrum by conditioning the input signal and transferring a downconverted version of it to a personal computer for software processing. Here, we employ a composite of two such dongles, interfaced with codes written in MATLAB and GNU Radio, as a measuring instrument to study the flexural vibrations of a few-layer graphene nanomechanical resonator. Instead of an antenna, we connect the dongles to the split output of a photodetector used to detect vibrations optically. We first perform a quantitative analysis of the dynamics of the first vibrational mode. We then measure the response of the first two vibrational modes in parallel. To illustrate our technique, we detect changes in the vibrational amplitude of both modes induced by periodic strain modulation with a delay of $\approx1$~ms between measurements. Last, we show that our software-based instrument can be employed to demodulate human voice encoded in the vibrations of our resonator. For parallel measurements of several frequency channels, and provided that the input signal is not too weak, our composite system may offer an alternative to the use of multiple lock-in amplifiers or multiple spectrum analyzers, with the distinct advantage of being cost-effective per frequency channel.
\end{abstract}

\maketitle

\section{Introduction}

Software-defined radio (SDR) has made the radio frequency spectrum accessible to everyone \cite{Molla2022}. Amateur radio enthusiasts employ SDRs as inexpensive receivers to demodulate signals they pick up from the air \cite{matlab}. Radio astronomers have developed large arrays of antennas where each antenna is connected to an SDR that dynamically amplifies and delays the received signal, making it possible to combine all signals in phase \cite{Dewdney2009,vanHaarlem2013,Bowman2018,Chime2018,Chime2020}. Electrical engineers and computer scientists have devised cognitive radio systems inspired by SDRs, where empty portions of the spectrum can be detected and dynamically allocated to communicate more efficiently \cite{Mitola1999,Fette2009}. The working principle of SDR is to perform as much processing of the radio signal as possible in software, leaving signal amplification and filtering to the analog front-end of the receiver. This technology was born of the wish for a single radio system that is agile, allowing the user to switch dynamically between demodulation modes \cite{Team1985,Mitola1992}. Some of the purest implementations of SDR, where the tasks of the analog front-end are minimal, are advanced systems developed by radio astronomers \cite{Lind2005}. Simpler and less expensive SDR devices, albeit with more limited performance, have become readily available in recent years. The latter, called SDR dongles, are the ones we use in this study.

SDR dongles evolved from tiny devices used to watch television on a computer \cite{matlab}. Those devices were based on an analog radio frequency (RF) tuner chip on the antenna side and a digital television demodulation chip on the computer side. Bypassing the demodulation step made it possible to pass frequency downconverted in-phase ($I$) and quadrature ($Q$) components of the received RF signal on to the computer, effectively turning the devices into inexpensive SDRs. Modern versions of such simple SDRs are the size of a USB flash drive \cite{RTLSDR}. They are valued by radio amateurs and scientists alike because of their wide frequency range, which extends from a few kilohertz to several gigahertz. They enable a variety of digital signal processing tasks that are challenging to perform with analog devices. These include processing signals with user-defined filters and dynamically changing demodulation modes, all of which are performed by the computer. Importantly, they come with open source software and drivers, enabling applications that are limited only by the imagination of the user. In a scientific context, SDR dongles have been used recently for oscillator metrology \cite{Sherman2016}, frequency modulation spectroscopy \cite{Mahnke2018}, optical interferometry \cite{Riobo2018}, magnetic resonance spectroscopy \cite{Doll2019}, and dual-comb spectroscopy \cite{Quevedo2021}. SDR dongles are also used as receivers in Radio Jove 2.0, a citizen science project supported by NASA whose participants measure RF noise originating from Jupiter and from the Sun \cite{RadioJove}.

Here, we demonstrate that SDR dongles provide an opportunity to facilitate the detection of nanomechanical vibrations. To the best of our knowledge, SDRs have not been employed to study nanomechanics thus far. Detecting the vibrations of nanomechanical resonators involves transducing their displacement amplitude into an electrical signal. Traditionally, this signal is further conditioned and processed with a lock-in amplifier or a spectrum analyzer. In our experimental setup, by contrast, signal conditioning and processing are performed with a composite of two SDR dongles running in parallel. The combination of (i) parallel SDRs and (ii) our data acquisition codes written in MATLAB and GNU Radio constitutes our measuring instrument. It allows us to easily measure two vibrational modes in parallel and to demodulate frequency-modulated vibrations.

The nanomechanical resonator we study is based on a suspended membrane of few-layer graphene (FLG) \cite{Bunch2007,Hone2009,Singh2010,vanderZande2010,Bouchiat2012,Song2012,Weber2014,Singh2014,Deshmukh2016,Davidovikj2016,DeAlba2016,Zhang2020}. FLG resonators have garnered much attention from scientists on account of their large in-plane stiffness \cite{Lee2008,Nicholl2015,Storch2018} and their small mass. Large in-plane stiffness causes resonant frequencies of flexural vibrations to range from a few megahertz to hundreds of megahertz, thereby spanning the high frequency and the very high frequency range of the radio spectrum. The small mass makes the resonators amenable to ultrasensitive detection schemes \cite{Review3}, including the detection of pressure changes \cite{Zhou2015,Dolleman2015,Wittman2019}, ultrasound \cite{Verbiest2018,Laitinen2019}, and light \cite{Blaikie2019}. These detection schemes are based on measuring amplitude changes and resonant frequency shifts of a single vibrational mode. In turn, measuring the response of several modes in parallel is advantageous, as it results in improved sensing accuracies. Such parallel measurements are the core principle of ultrasensitive mass sensing experiments based on resonators fabricated from bulk materials \cite{Roukes2012,Roukes2015,Hentz2015}. They require one rather advanced measuring instrument, such as a lock-in amplifier or a spectrum analyzer, per vibrational mode. Our work explores an alternate approach based on a composite of two SDR dongles. We measure the frequency response of the first two vibrational modes of our FLG resonator in parallel with a delay of $\approx1$~ms between the two measurements. This is useful to verify whether any anomalous feature in the lineshape of one mode at a given time is also observed in the lineshape of the other mode almost at the same time, indicating, e.g., changes in mass or strain. To substantiate this idea, we modulate strain within the resonator with steps of dc voltage applied between the membrane and an underlying gate electrode, and simultaneously record changes in the amplitude of the modes at two off-resonance frequencies. Finally, to complement our software-based measurement approach, we demonstrate demodulation of frequency-modulated driven vibrations in software, using human voice and musical waveforms to modulate the driving force.

We believe that the advantages of our approach are twofold. First, our technique allows us to monitor two channels far apart in frequency with only a short time delay. As mentioned above, the same may alternatively be achieved with two lock-in amplifiers or two spectrum analyzers. However, given that the input signal gets divided among SDRs, using multiple SDRs suddenly makes our system quite inexpensive per channel. Second, the fact that SDR outputs $I$ and $Q$ components of the input signal means that any type of demodulation can be performed dynamically in software. This is beneficial for decoding any type of information imprinted in vibrations \cite{Hone2013}. Overall, we show that employing SDR dongles in a vibration detection setup is a simple yet useful and versatile method to acquire and process nanomechanical signals.

\section{Experimental setup}

Below, we briefly describe the structure of our resonator and present our vibration detection setup. We summarize the process by which SDRs receive and condition the nanomechanical signal. Then, we characterize the intrinsic frequency stability of the signal at the output of one of our SDRs using a low phase noise input signal.
\begin{figure*}[t]
\includegraphics{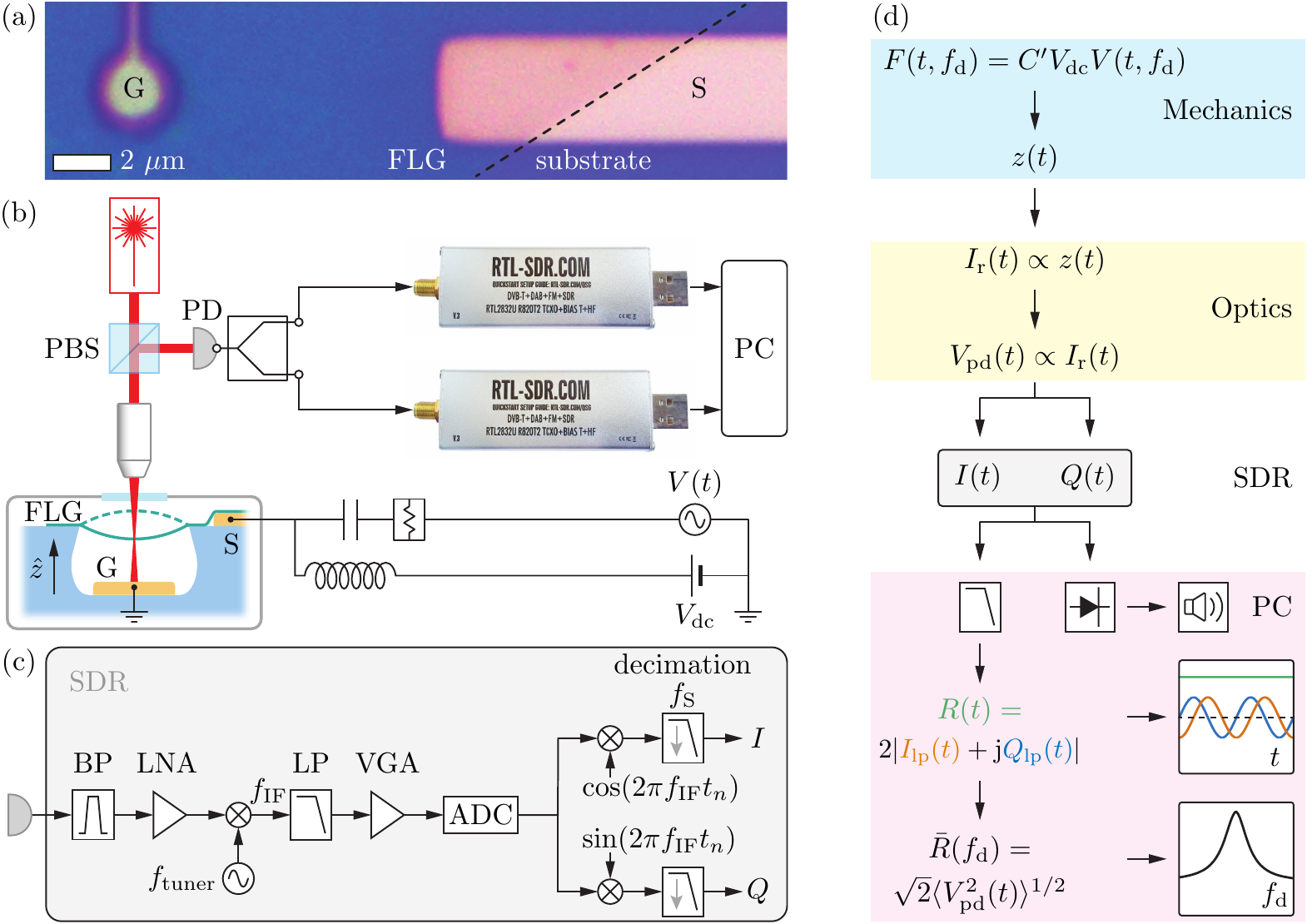}
\caption{Resonator and measurement setup. (a) Optical micrograph of our few-layer graphene nanomechanical resonator. G, S: gate and source electrodes. FLG: few-layer graphene. (b) Setup. Output of photodetector (PD) is split and fed to two SDR dongles. PBS: polarizing beam splitter. (c) Simplified block diagram for a single SDR. BP: bandpass filter. LNA: low noise amplifier. $f_\mathrm{tuner}$: frequency of local oscillator. $f_\mathrm{IF}$: intermediary frequency. VGA: variable gain amplifier. ADC: analog-to-digital converter. Downstream from ADC, in-phase and quadrature components of the digitized signal are downconverted to $I$ and $Q$ baseband using two mixers and two decimation filters with sampling rate $f_\mathrm{S}$. Discrete time $t_n=n/f_\mathrm{ADC}$, with $f_\mathrm{ADC}$ the ADC sampling rate and $n$ spanning a range of integers. (d) Flowchart depicting our measurement process from the driving force $F$ to the measured signal in the time domain $R(t)$ and in the frequency domain $\bar{R}(f_\mathrm{d})$. The diode symbol represents demodulation in software.}\label{fig1}
\end{figure*}

Our resonator is shown in Fig.~\ref{fig1}a. It is fabricated by transferring a membrane of FLG onto a substrate that we previously patterned using electron beam lithography \cite{Transfer}. The substrate is made of highly resistive silicon overgrown with thermal silicon oxide. A cylindrical cavity, 250~nm deep and 3.1~$\mu$m in diameter, is etched in the oxide layer. Metal electrodes are fabricated both on the top surface of the substrate and at the bottom of the cavity (Figs.~\ref{fig1}a, b). The electrode on the top surface is used to electrically contact the membrane, and the electrode at the bottom of the cavity serves as a gate electrode. Together, the membrane and the gate electrode form a capacitor whose capacitance $C$, according to our COMSOL simulations, is well approximated by that of a regular parallel plate capacitor, even where the membrane is slightly bent towards the gate. This capacitor configuration enables driving vibrations electrically \cite{Sazanova}. Coherent drive is achieved by applying a dc voltage $V_\mathrm{dc}$ and an oscillating voltage of peak amplitude $V$ and frequency $f_\mathrm{d}$ between the membrane and the gate. The resulting driving force at $f_\mathrm{d}$ has a peak amplitude $\approx C^\prime V_\mathrm{dc}V$, with $C^\prime=\mathrm{d}C/\mathrm{d}z$ being the derivative of $C$ with respect to flexural displacement in the direction $\hat{z}$ (Fig.~\ref{fig1}b). Bandpass filtering of the output of the signal generator that delivers the oscillating voltage ensures that harmonics and spurs below and above $f_\mathrm{d}$ are attenuated. The resonator is kept at room temperature in a vacuum of $\approx10^{-6}$~mbar \cite{Heng}.

Flexural vibrations are measured optically using a setup first reported in \cite{Carr1997,Erkinci2005} and first employed with FLG resonators in \cite{Bunch2007}. Briefly, the resonator is placed in an optical standing wave from which it periodically absorbs energy as it vibrates \cite{Barton2012}. The standing wave is formed by the superposition of an incident wave originating from a helium-neon laser and the wave reflected by the gate electrode acting as a mirror. As a function of time $t$, the intensity of reflected light $I_\mathrm{r}(t)$ is modulated near the resonant frequency of vibrations by the amount of energy that the membrane periodically absorbs \cite{Barton2012,Heng}. For a fully reflective gate electrode, the amplitude of this modulation reads $I_\mathrm{r}(t)\approx I_\mathrm{inc}[1-A-(\mathrm{d}A/\mathrm{d}z)z(t)]$, where $I_\mathrm{inc}$ is the intensity of light incident on the resonator, $\mathrm{d}A/\mathrm{d}z\approx-6\times10^{-3}$~nm$^{-1}$ is the derivative of the absorbance $A$ of the membrane with respect to displacement \cite{Roddaro2007,Fengnan}, and $z(t)=z_0(f_\mathrm{d})\cos[2\pi f_\mathrm{d}t+\phi(f_\mathrm{d})]$ is the displacement of the membrane in the flexural direction with vibrational amplitude $z_0$ and phase $\phi$ with respect to the driving force at frequency $f_\mathrm{d}$. We measure $I_\mathrm{r}(t)$ with an avalanche photodetector, whose output, after high-pass filtering of the dc component, is a radio frequency voltage $V_\mathrm{pd}(t)\propto I_\mathrm{inc}(\mathrm{d}A/\mathrm{d}z)z(t)$.

We measure $V_\mathrm{pd}(t)$ with SDR dongles. We have used both RTL-SDR Blog V3 by RTL-SDR Blog and NESDR Smart V4 by Nooelec \cite{RTLSDR}, and obtained quantitatively similar results. To explain our measurement protocol, we find it helpful to first briefly describe the basic principle of operation of these SDR dongles (Fig.~\ref{fig1}c and~\cite{matlab}). The front-end of a dongle acts as a heterodyne receiver. The analog RF signal from the photodetector output is amplified with a low noise amplifier. The amplified signal is fed to a tuner that selects a band about 2.8~MHz wide and downconverts it to an intermediary frequency (IF) $f_\mathrm{IF}=3.57$~MHz. This downconversion is performed with an analog mixer and a voltage-controlled oscillator of tunable frequency $f_\mathrm{tuner}$. The downconverted frequency band centered at $f_\mathrm{IF}$ is further low-pass filtered and amplified and then digitized at a rate $f_\mathrm{ADC}=28.8$~MHz. The digitized IF signal is fed to a complex digital mixer which downconverts it into a complex baseband signal near 0~Hz. The $I$ and $Q$ components of this complex baseband signal are important because they help preserve the integrity of the downconverted spectrum, which has both positive and negative frequency components. The baseband signal is low-pass filtered to reduce its bandwidth and then decimated, allowing it to be resampled at a rate $f_\mathrm{S}$ that is much lower than $f_\mathrm{ADC}$. This decimated baseband signal is sent to the computer. In its simplest form, and in the case of a voltage amplification factor of 1, SDR transforms an input signal $H(t)=H\cos(2\pi f_\mathrm{d}t+\phi)$ into $I(t)=\frac{H}{2}\cos(2\pi\Delta ft+\phi)$ and $Q(t)=-\frac{H}{2}\sin(2\pi\Delta ft+\phi)$, with $\Delta f= |f_\mathrm{tuner}+f_\mathrm{IF}-f_\mathrm{d}|$. For clarity, our full measurement process, from the transduction of a driving force into a resonant displacement all the way to our data processing in software, is schematically depicted in Fig.~\ref{fig1}d.

\begin{figure}[t]
\includegraphics{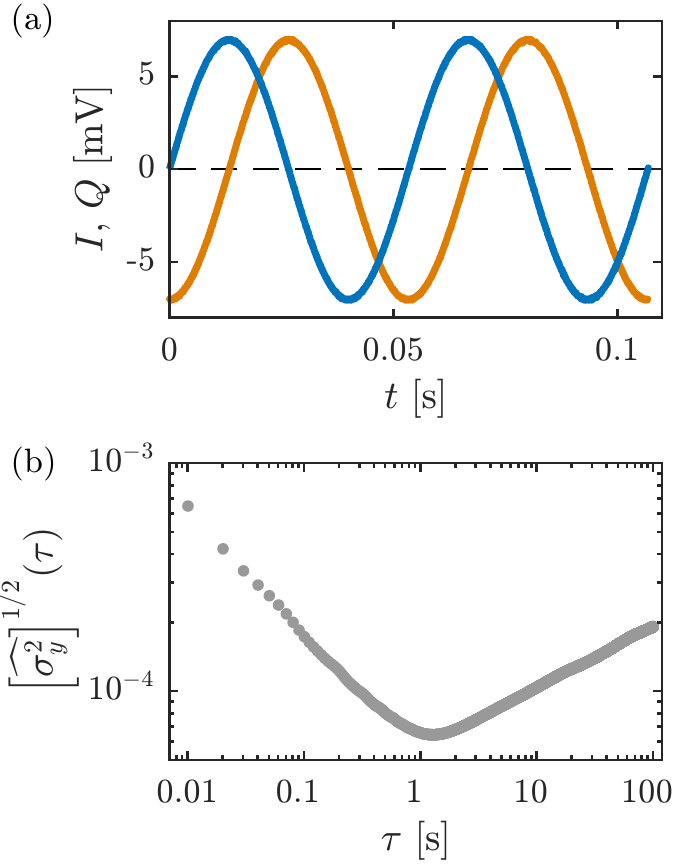}
\caption{Characterizing the frequency stability of SDR's output signal. (a) In-phase $I$ and quadrature $Q$ components of the voltage signal at the output of SDR as a function of time $t$ for a harmonic input voltage at frequency $f_\mathrm{in}=41$~MHz. $\Delta f=|f_\mathrm{tuner}+f_\mathrm{IF}-f_\mathrm{in}|\approx20$~Hz. (b) Estimate of the Allan deviation of fractional frequencies of the output signal as a function of integration time $\tau$ for $\Delta f\approx100$~Hz.}\label{fig2}
\end{figure}

Prior to measuring the response of our graphene resonator, we characterize the intrinsic frequency stability of the output signal of one of our SDRs. We estimate this stability from the Allan variance of fractional frequency fluctuations in the output signal. To the best of our knowledge, this characterization has not been reported thus far. However, it is needed for resonant detection schemes based on SDR. Fig.~\ref{fig2}a shows $I(t)$ and $Q(t)$ components of the decimated baseband signal in the simple case where a harmonic input signal at frequency $f_\mathrm{in}=41$~MHz is applied to the SDR. The input signal is supplied by a low amplitude noise, low phase noise generator (N5181B by Keysight). We have previously calibrated the output amplitude of the SDR using the signal generator as an input and verified that the SDR built-in amplifiers are operating in their linear regime. We have also measured $I$ and $Q$ with a 50~Ohm matched attenuator at the SDR input to prevent standing waves from building up between the generator and SDR, and have not observed any change in the quality of the output signal. We acquire data with our MATLAB code based on open source drivers \cite{OSMOCOM}. Defining $\Delta f=|f_\mathrm{tuner}+f_\mathrm{IF}-f_\mathrm{in}|$, we set $\Delta f=20$~Hz and sampling rate $f_\mathrm{S}=2.5\times10^5$~S/s (samples per second). $I(t)$ and $Q(t)$ oscillate harmonically at frequency $\approx20$~Hz and are in quadrature, as expected. We then set $\Delta f=100$~Hz (the actual oscillation frequency is off by a few Hz on average) and continuously sample $I$ and $Q$. Using decimation at $f_\mathrm{S}$ as our time base and detecting zero crossing times in $I(t)$, we build an array of instantaneous periods $T_p$ with $p\in\mathbb{N}$. From this array, we build the time-error function $x(n\tau_0)\equiv x_n=\sum_{p=1}^nT_p-n\tau_0$, with $n\in\mathbb{N}$ and where $\tau_0=\langle T_p\rangle$ is the ensemble average of $T_p$ (we find that $\tau_0\approx1/\Delta f$). We then compute an experimental estimate for the Allan variance of fractional frequencies $y$ of $I(t)$, see \cite{Howe1981}:
\begin{equation}
\begin{aligned}
\widehat{\sigma_y^2}(m\tau_0)&=\frac{1}{2(N-2m)m^2\tau_0^2}\\
&\times\sum_{i=1}^{N-2m}\left(x_{i+2m}-2x_{i+m}+x_i\right)^2\,,\label{AVAR}
\end{aligned}
\end{equation}
where $m$ is an integer, $N$ is the number of terms in the sequence $(x_1,x_2,x_3,\ldots)$, and $m\tau_0$ is an integration time. The estimate of the Allan deviation, $\Big[\widehat{\sigma_y^2}\Big]^{1/2}(\tau)$, with $\tau=m\tau_0$, is shown in Fig.~\ref{fig2}b. It exhibits a minimum $\sigma_y^\mathrm{min}\approx6\times10^{-5}$ near $\tau\approx1$~s. This minimum indicates that the root-mean-square deviation $\delta f$ between two measurements of the frequency of $I(t)$, made $\approx1$~s apart, is $\delta f=\sigma_y^\mathrm{min}\langle1/T_p\rangle\approx\sigma_y^\mathrm{min}/\langle T_p\rangle\approx\sigma_y^\mathrm{min}\Delta f\approx6\times10^{-3}$~Hz. If we assume that these frequency fluctuations originate from the SDR oscillator, then they are translated from $f_\mathrm{in}=41$~MHz down to $\Delta f$. In this case, the frequency stability of the oscillator is $\delta f/f_\mathrm{in}\approx1.5\times10^{-10}$. This value is small, yet it is still $\approx300$ times larger than the frequency stability set by frequency flicker noise in the input signal which we estimate from the phase noise of the signal generator (Appendix~A). However, our estimate of $\delta f/f_\mathrm{in}$ is at least one order of magnitude smaller than $\sigma_y^\mathrm{min}$ in nanomechanical resonators \cite{Hentz2016}, making SDR suitable for performing resonant measurements with these systems.

\section{Measuring flexural vibrations with SDR}

\begin{figure*}[t]
\includegraphics{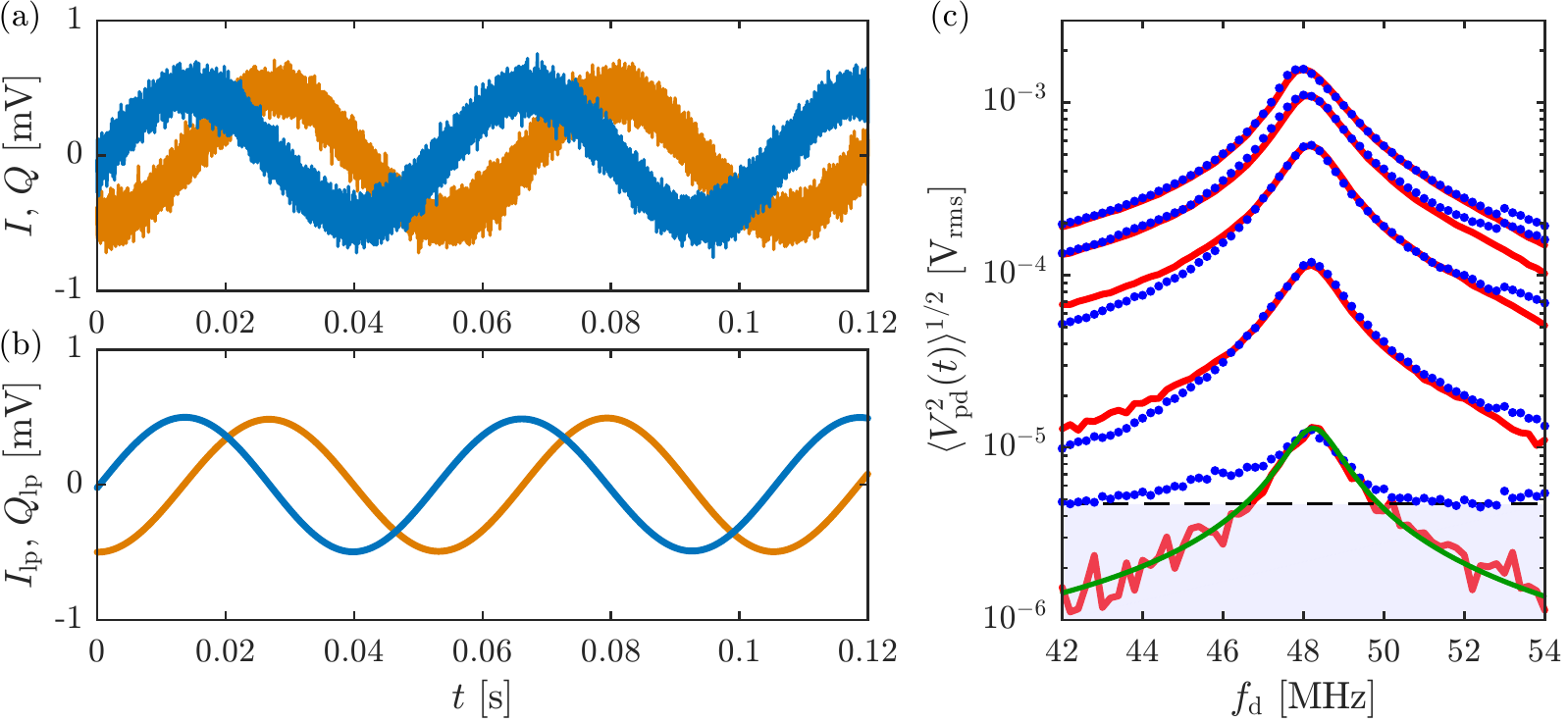}
\caption{Measuring the frequency response of the first vibrational mode of our FLG resonator. (a) Typical $I(t)$ and $Q(t)$ traces obtained by driving the resonator near resonance. Sampling rate $f_\mathrm{S}=2.5\times10^5$~S/s. (b) Low-pass filtered traces, $I_\mathrm{lp}(t)$ and $Q_\mathrm{lp}(t)$, obtained by processing traces in (a) with a Butterworth filter in software. (c) Response of the resonator measured with SDR (blue dotted traces) and with a low noise spectrum analyzer (red traces). From bottom to top: $V=0.22$, $2.24$, $11.19$, $22.39$, and $33.58$~mV, where $V$ is the peak amplitude of the driving voltage (accounting for full reflection of voltage waves incident on the resonator due to impedance mismatch). $f_\mathrm{S}=2.5\times10^5$~S/s. Both SDR and spectrum analyzer data are averaged for 1~s. Green trace: fit to~\eqref{eqn} for $V=0.22$~mV. Horizontal dashed line and gray shading indicate the measurement background of the SDR.}\label{fig3}
\end{figure*}
Having characterized the frequency stability of SDR, we measure the frequency response of our FLG resonator. Fig.~\ref{fig1}b shows that the output of the photodetector is split into two paths, each of which is connected to the SMA (SubMiniature version A) port of one SDR. We use one of these to measure the response of the first (lowest frequency) vibrational mode. Fig.~\ref{fig3}a shows typical $I(t)$ and $Q(t)$ baseband traces obtained with the driving frequency $f_\mathrm{d}$ near resonance and $f_\mathrm{S}=2.5\times10^5$~S/s. We filter these traces in software with low-pass Butterworth filters and obtain $I_\mathrm{lp}(t)$ and $Q_\mathrm{lp}(t)$ (Fig.~\ref{fig3}b). At fixed driving voltage $V$, we acquire $I_\mathrm{lp}(t)$ and $Q_\mathrm{lp}(t)$ for 1~s and compute $R(t)=2|I_\mathrm{lp}(t)+\mathrm{j}Q_\mathrm{lp}(t)|$, which corresponds to the peak amplitude of the output of the photodetector $V_\mathrm{pd}(t)$. Because of weak amplitude fluctuations in $I_\mathrm{lp}(t)$ and $Q_\mathrm{lp}(t)$, $R(t)$ is weakly time dependent. We average $R(t)$ over time to obtain an estimate $\bar{R}$ of the amplitude of $V_\mathrm{pd}(t)$. Finally, we estimate the root-mean-square amplitude $\langle V_\mathrm{pd}^2(t)\rangle^{1/2}\approx\bar{R}/\sqrt{2}$, where $\langle\cdot\rangle$ is an average over many mechanical periods. Fig.~\ref{fig3}c displays the frequency response $\langle V_\mathrm{pd}^2(t)\rangle^{1/2}(f_\mathrm{d})$ of our resonator for $V$ ranging from 0.22~mV to 33.58~mV (peak amplitude) and at $V_\mathrm{dc}=11$~V, see blue dotted traces (the response at $V_\mathrm{dc}=5$~V is shown in Appendix~B. Also shown is the response measured with a low noise spectrum analyzer (FSW8 by Rohde \& Schwarz) under the same conditions and some time thereafter (red traces). Because the traces in blue are obtained using various gains of the SDR amplifiers, we scale their peak amplitude to match the peak amplitude of the red traces. (This amplitude calibration is consistent with an alternate calibration we obtain using a signal of known amplitude at the input of SDR.) The two sets of measurements depicted by the blue and the red traces are reasonably consistent with each other. At $V=0.22$~mV and away from resonance, $\langle V_\mathrm{pd}^2(t)\rangle^{1/2}(f_\mathrm{d})$ is limited by the measurement background of the SDR (gray shading bounded by the horizontal dashed line), whereas the spectrum analyzer can still resolve the response. This measurement yields the weakest electromechanical signal $\approx5\times10^{-6}$~V$_\mathrm{rms}$ our SDR can resolve.

\begin{figure*}[t]
\includegraphics{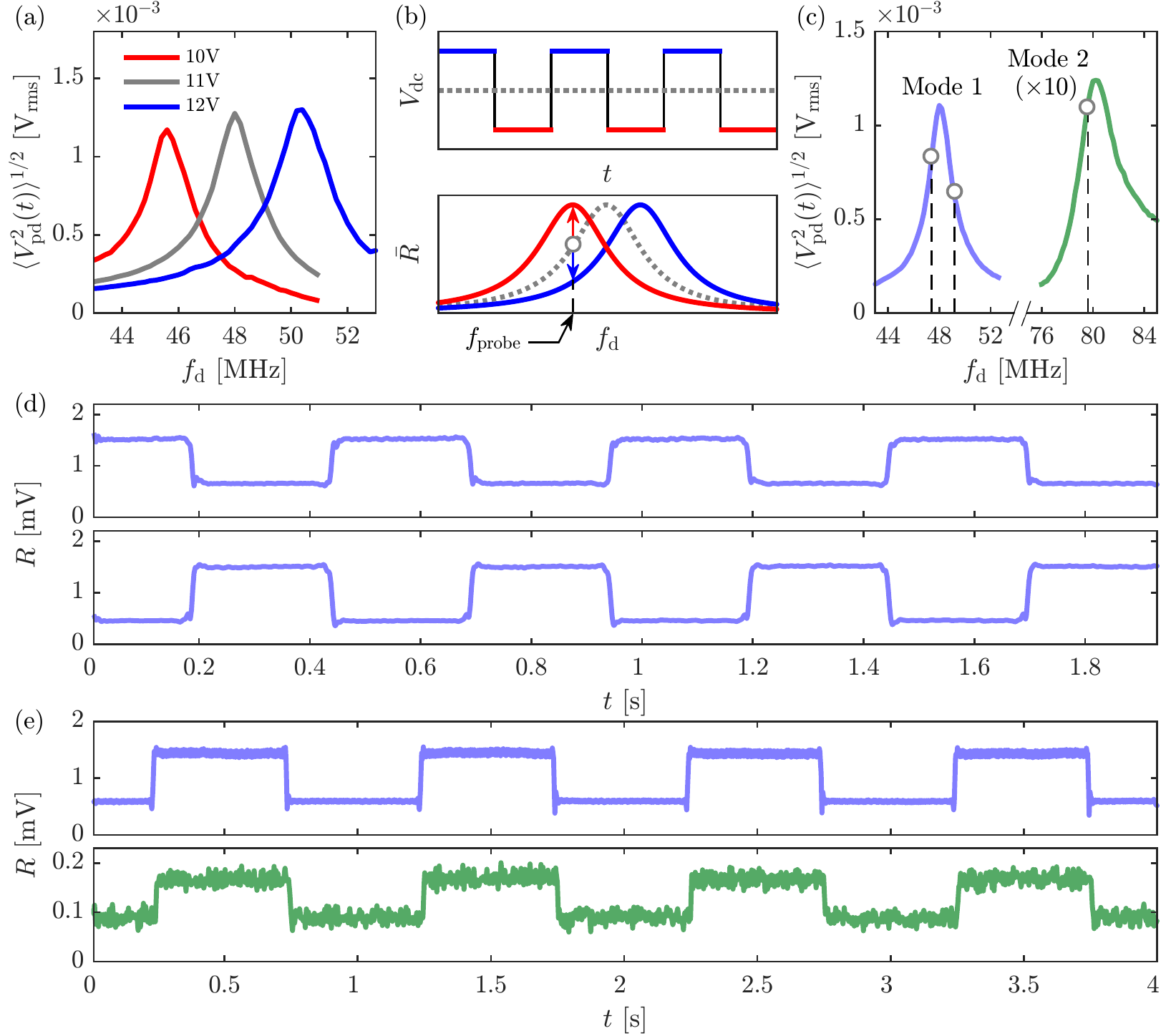}
\caption{Measuring two vibrational modes in parallel with SDR. (a) Response of the first vibrational mode measured at $V_\mathrm{dc}=10$, $11$, and $12$~V. (b) Schematic of the strain modulation experiment. For $V_\mathrm{dc}$ below (resp. above) the gray dotted line in the upper panel, the response in the lower panel shifts to lower (resp. higher) frequency and $\bar{R}(f_\mathrm{probe})$ increases (resp. decreases). (c) Responses of first and second vibrational modes (Mode~1 and Mode~2) measured in parallel ($V_\mathrm{dc}=11$~V). (d) Response of Mode~1 with $V_\mathrm{dc}$ modulated by $\pm0.35$~V above and below 11~V. The upper (resp. lower) panel shows $R(t)$ measured with driving frequency $f_\mathrm{d}=f_\mathrm{probe}$ set below (resp. above) the resonant frequency of Mode~1 at $V_\mathrm{dc}=11$~V with both drives on (vertical dashed lines in (c)). (e) Responses of Mode~1 (upper panel) and Mode~2 (lower panel) with $V_\mathrm{dc}$ modulated by $\pm0.5$~V above and below 11~V. Drive frequencies are set below resonance (panel c). Peak voltage amplitude of all drives incident on the resonator is $V=22.4$~mV in all panels. Sampling rate $f_\mathrm{S}=2.5\times10^5$~S/s in (a), (c), and (e); $f_\mathrm{S}=2.048\times10^6$~S/s in (d).}\label{fig4}
\end{figure*}
We use the measured response to quantify the dynamics of our resonator. For this, we fit $\langle V_\mathrm{pd}^2(t)\rangle^{1/2}(f_\mathrm{d})\propto\langle z^2(t)\rangle^{1/2}(f_\mathrm{d})$ to the standard model of a single harmonic resonator subjected to a driving force $F\cos(2\pi f_\mathrm{d}t)$, to a nonlinear damping force \cite{Dykman1975,Buks2012} of the form $[\gamma+\eta z^2(t)]\mathrm{d}z/\mathrm{d}t$, and to a nonlinear restoring force $[k_0+\alpha z^2(t)]z(t)$. Linear and nonlinear damping coefficients are $\gamma$ and $\eta$, respectively. The linear spring constant is $k_0=M\omega_0^2$, with $M$ the effective mass of the vibrational mode and $\omega_0=2\pi f_0$ its angular resonant frequency, and the Duffing parameter of the nonlinear restoring force is $\alpha$. Using the method of averaging, we write the equation of motion of the harmonic resonator as follows:
\begin{equation}
\begin{aligned}
&\xi^3\left[\frac{9}{16}\tilde{\alpha}^2+\frac{1}{16}\tilde{\eta}^2\omega^2\right]+\xi^2\left[-\frac{3}{2}\tilde{\alpha}\left(\omega^2-\omega_0^2\right)+\frac{1}{2}\tilde{\eta}\tilde{\gamma}\omega^2\right]\\
&+\xi\left[\left(\omega^2-\omega_0^2\right)^2+\tilde{\gamma}^2\omega^2\right]-\tilde{F}^2=0\,,\label{eqn}
\end{aligned}
\end{equation}
where $\xi=\bar{z_0}^2$ with $\bar{z_0}$ the time-averaged vibrational amplitude (which is positive real), and where $\tilde{\alpha}=\alpha/M$, $\tilde{\eta}=\eta/M$, $\tilde{\gamma}=\gamma/M$, $\tilde{F}=F/M$, and $\omega=2\pi f_\mathrm{d}$. The effective mass is estimated by combining the number of layers $N_\mathrm{L}=15$ in the membrane measured as in \cite{Heng}, the diameter $\approx3.1$~$\mu$m of the suspended part of the membrane, and the vibrational mode shape measured as in \cite{Heng}, yielding $M\approx2.3\times10^{-17}$~kg. The derivative of the capacitance $C^\prime\approx4\times10^{-10}$~F/m is obtained with COMSOL. The green trace in Fig.~\ref{fig3}c is a fit to~\eqref{eqn} of the response measured with the spectrum analyzer using the lowest drive. For clarity, responses measured with larger drives are fit to~\eqref{eqn} in Appendix~B. This analysis yields $\alpha\approx-(5\pm1)\times10^{15}$~kg/(m$^2\cdot$s$^2$), $\eta\approx(5\pm2)\times10^6$~kg/(m$^2\cdot$s) and $\gamma\approx(1.9\pm0.1)\times10^{-10}$~kg/s. Nonlinearities arising from external potentials may account for $\alpha<0$ \cite{Cross_Lifshitz2008}. Overall, we show that SDR can be used to reliably measure the response of the first vibrational mode and to characterize its dynamics.

The true advantage of SDR technology may be best appreciated in the case of a composite of multiple calibrated SDR dongles. For example, such a composite would be useful in a mass sensing experiment. There, monitoring the amplitude of at least two vibrational modes in parallel, that is, measuring multiple modes `at the same time', makes it possible to identify both the mass of the adsorbate and the location of the adsorption site on the resonator \cite{Roukes2015}. In those experiments \cite{Roukes2012,Hentz2015}, two modes are measured using two lock-in amplifiers, enabling low noise vector measurements of the modal responses. If, instead, it is sufficient to measure the magnitude of the responses, then a composite of SDR dongles offers an alternative that is remarkably simple and inexpensive per frequency channel. Here, we justify this idea not by adding mass to the resonator, but by modulating strain within the resonator and monitoring changes in response amplitude. Fig.~\ref{fig4}a shows the frequency response of the first mode measured with SDR at 3 different gate voltages $V_\mathrm{dc}$. The resonant frequency $f_0$ changes by $\mathrm{d}f_0/\mathrm{d}V_\mathrm{dc}\approx2.4$~MHz/V due to stretching of the membrane by the electrostatic force. This suggests the experiment schematically depicted in Fig.~\ref{fig4}b, where we set the driving frequency $f_\mathrm{d}=f_\mathrm{probe}$ slightly off resonance for a given value of $V_\mathrm{dc}$ (gray dashed line), modulate $V_\mathrm{dc}$ between two levels above and below this value (upper panel, blue and red lines), and expect resulting modulations of $R(t)$ at $f_\mathrm{probe}$ (lower panel). For this experiment, we split the output of the photodetector (Fig.~\ref{fig1}b), connect each split path to one SDR dongle, and connect the two dongles to a single computer. Each dongle acts a server that streams data using the Transmission Control Protocol; it is controlled by our MATLAB code (Appendix~C). To probe the linear response at two different frequencies $f_\mathrm{d}^\prime$ and $f_\mathrm{d}^{\prime\prime}$, we use two synchronized signal generators and make the carrier frequency of each one equal to the center frequency $f_\mathrm{tuner}+f_\mathrm{IF}$ of one dongle. At both $f_\mathrm{d}^\prime$ and $f_\mathrm{d}^{\prime\prime}$, driving voltages incident on the resonator have the same peak amplitude $V=22.4$~mV.

We first measure the response of the first vibrational mode near 48~MHz (Mode~1) and that of the second mode near 80~MHz (Mode~2) in parallel at $V_\mathrm{dc}=11$~V. Stepping $f_\mathrm{d}^\prime$ and $f_\mathrm{d}^{\prime\prime}$ together, sampling each mode at $f_\mathrm{S}=2.5\times10^5$~S/s and averaging $R(t)$ measured by each dongle for 1~s yields the responses shown in Fig.~\ref{fig4}c. For each pair $(f_\mathrm{d}^\prime,f_\mathrm{d}^{\prime\prime})$, the signal from Mode~1 and that from Mode~2 are received together. This technique has allowed us to identify occasional mechanical instabilities in the resonator, whereby sudden changes in amplitude would be observed in both modes at the same time. If the two modes are measured separately in time, one measured response may appear to be anomalous while the other may appear as regular, even though both responses have been affected by the instability.

We then measure the responses in the presence of modulated $V_\mathrm{dc}$. In Fig.~\ref{fig4}d, we set $f_\mathrm{d}^\prime$ below resonance of Mode~1 and $f_\mathrm{d}^{\prime\prime}$ above resonance of Mode~1 at $V_\mathrm{dc}=11$~V, both within the bandwidth of the mode (pair of vertical dashed lines in panel c). We then modulate $V_\mathrm{dc}$ periodically by $\Delta V_\mathrm{dc}=\pm0.35$~V above and below 11~V. This results in a modulation of the resonant frequency $\Delta f_0=\Delta V_\mathrm{dc}\mathrm{d}f_0/\mathrm{d}V_\mathrm{dc}=\pm0.84\times10^6$~Hz that periodically shifts the response in frequency. Correspondingly, $R(t)$ measured at each driving frequency is amplitude modulated by an amount $\Delta R\approx\Delta f_0\times\mathrm{d}R/\mathrm{d}f_\mathrm{d}$. Because the response is probed below and above resonance, the two traces in panel (d) are anticorrelated. It is also interesting to estimate how `simultaneous' these time traces are. We find that maxima in $|\mathrm{d}R/\mathrm{d}t|$ for the upper trace and those for the lower trace occur within $\approx1$~ms of each other (Appendix~D). This time scale is also the rise time of $V_\mathrm{dc}$ steps at the output of the bias tee in Fig.~\ref{fig1}b. It would be sufficiently short for mass sensing experiments in vacuum, where adsorption events have been shown to occur ever few seconds \cite{Chaste2012}. Fig.~\ref{fig4}e shows the response of Mode~1 (upper panel) and that of Mode~2 (lower panel). $f_\mathrm{d}^{\prime}$ is set below resonance of Mode~1, $f_\mathrm{d}^{\prime\prime}$ is set below resonance of Mode~2 (vertical dashed lines below 48~MHz and below 80~MHz in panel c), and $\Delta V_\mathrm{dc}=\pm0.5$~V (near $V_\mathrm{dc}=11$~V, $\mathrm{d}f_1/\mathrm{d}V_\mathrm{dc}\approx1.6$~MHz/V, where $f_1$ is the resonant frequency of Mode~2: see Appendix~E). Time traces in panel (e) clearly show that the two modes are simultaneously modified by a change in strain (the two traces are correlated). Overall, panel (e) substantiates the idea that  multiple modes can be measured in parallel with a composite of SDR dongles, provided that the input signal is large enough that it can be split among the dongles.

\begin{figure*}[t]
\includegraphics{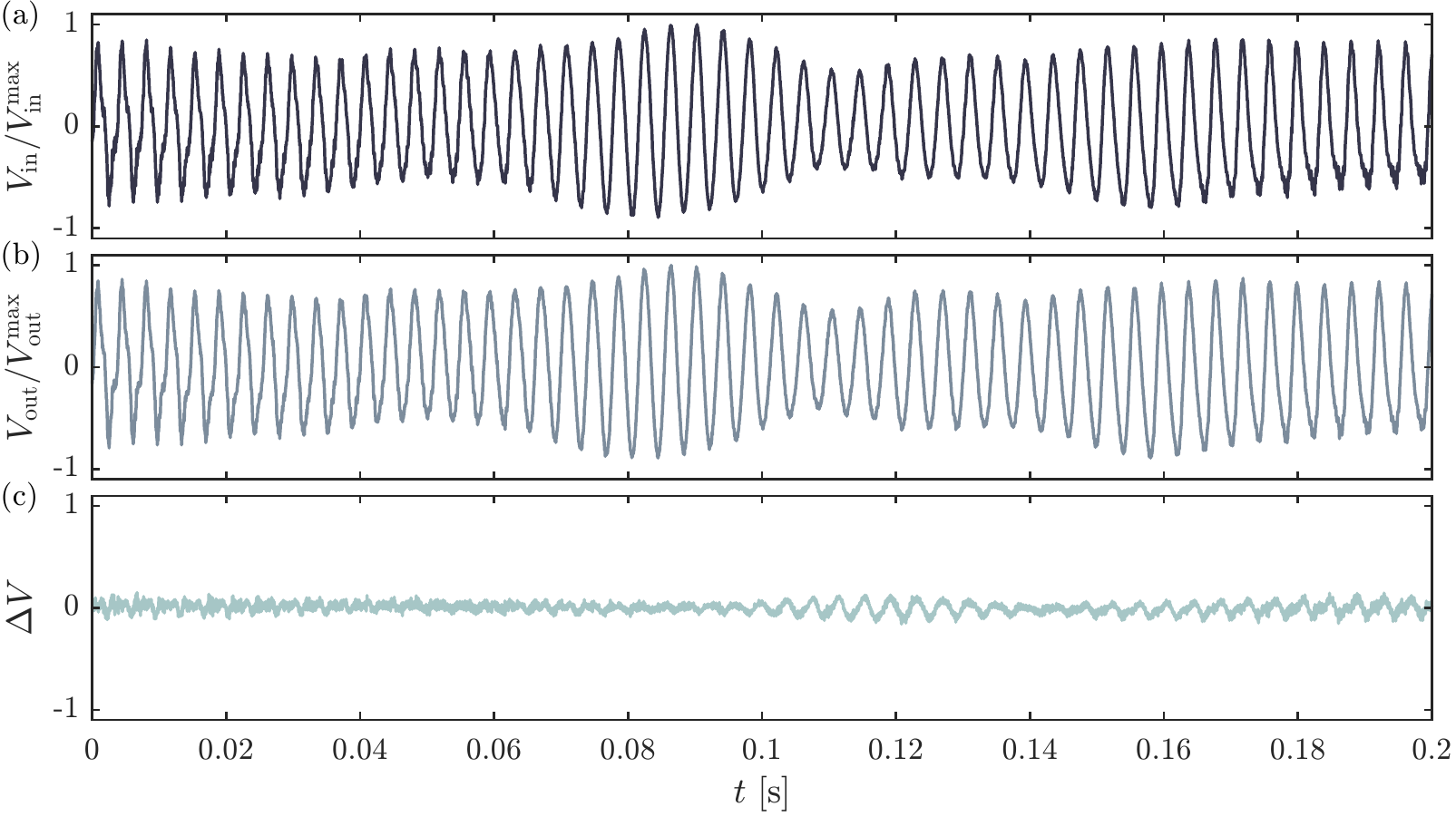}
\caption{Extracting information encoded in the optical field by vibrations. (a) Excerpt from the normalized waveform data of the original Taihu Mei audio recording. The file in mp3 format can be heard in Appendix~G. (b) Corresponding excerpt from the normalized, demodulated signal at the output of SDR. The demodulated signal can be heard in Appendix~G. (c) Difference $\Delta V=V_\mathrm{out}/\mathrm{max}(V_\mathrm{out})-V_\mathrm{in}/\mathrm{max}(V_\mathrm{in})$, showing that modulations in the drive are properly recovered after demodulation of the measured response.}\label{fig5}
\end{figure*}
Finally, to complement our software-based measurement approach, we demonstrate that SDR can be employed to extract information encoded in the optical field by the vibrations of our resonator. Here, information encoding is performed by driving the resonator with a frequency modulated (FM) voltage waveform $V_\mathrm{in}(t)$ whose instantaneous frequency $f_\mathrm{inst}$ is time dependent. The latter reads $f_\mathrm{inst}(t)=f_0+f_\Delta u(t)$, where $f_\Delta$ is the frequency deviation and $u(t)$ is the baseband signal to be encoded. To create $u(t)$, we use a digital audio recording of a song performed by one of the authors (J. Wang). A similar experiment was reported in \cite{Hone2013}, with three differences: information was encoded into an electromechanical signal instead of an optomechanical signal; an analog hardware demodulator was used instead of SDR and software; pop music was played instead of an unprocessed human voice. Our recording is fed from the output of the computer sound card to the input of a voltage-controlled oscillator used as a baseband frequency modulator in our signal generator. We use a driving voltage of peak amplitude $V=25$~mV that is low enough to keep the resonator in its linear regime and large enough to obtain a good signal-to-noise ratio at the SDR output. We set $f_\Delta=40$~kHz, which is large enough to encode most of the frequency components of the audio signal and yet much smaller than the linewidth of the mechanical response. We set $f_\mathrm{tuner}+f_\mathrm{IF}$ close to the resonance of the first mode. On the computer, we digitally filter $I$ and $Q$ with a 100~kHz-wide bandpass filter and digitally demodulate them to extract the audio baseband with our GNU Radio code (Appendix~F). Fig.~\ref{fig5}a displays an excerpt from $V_\mathrm{in}(t)$, normalized to its maximum amplitude, taken from the original audio recording. Fig.~\ref{fig5}b represents the corresponding excerpt $V_\mathrm{out}(t)$, normalized to its maximum amplitude, extracted from the demodulated signal. Fig.~\ref{fig5}c shows the difference between the two waveforms. The latter is small, indicating that the baseband signal encoded in the mechanical vibrations is faithfully imprinted in the intensity of reflected light and is properly recovered with our software-based signal processing technique. The original recording in mp3 format can be heard in Appendix~G. The song, entitled ``Taihu Mei'' (meaning ``Beautiful Lake Tai''), is performed in a Chinese dialect spoken in the city of Suzhou. It is derived from a traditional ditty named ``Wuxi Jing'' that was composed at the end of the Qing Dynasty. The demodulated signal in mp3 format can also be heard in Appendix~G. More of our demodulated nanomechanical signals, including poetry by Shakespeare and an excerpt from a musical, can be heard in Appendix~G.

\section{Limitations of the study}

Having shown that our composite of SDRs can be used to measure the driven vibrations of a few-layer graphene resonator at room temperature, we now address the suitability of our approach for measuring them at low temperature. The dynamics of these resonators at low temperature is interesting in part because their quality factor $Q$ increases as temperature decreases \cite{Hone2009}. This is advantageous to resonant sensing. For example, the smallest mass and the smallest force detectable in a resonant experiment are both proportional to $Q^{-1/2}$. However, measuring vibrations at low temperature is challenging because the probe signal must be weak enough for the resonator not to heat up. Many of those low temperature measurements reported thus far are based on an electrical technique first employed at room temperature in~\cite{Sazanova}. There, the probe signal is an oscillating voltage $V_\mathrm{sd}(t)$ applied between the source and drain electrodes of the resonator, which takes the role of the incident optical wave in our study. The magnitude of the measured mechanical response is proportional to $|V_\mathrm{sd}(t)|$. In \cite{Moser2014}, the weakest driven vibrational amplitude that can be resolved at 10~mK with this technique is measured as a voltage of amplitude $\approx1$~nV$_\mathrm{rms}$ (Fig.~1c in \cite{Moser2014}, where an electromechanical current of smallest amplitude $\approx1$~pA is converted into a voltage using a 2~kOhm resistor). With regard to our Fig.~\ref{fig3}c, 1~nV$_\mathrm{rms}$ is 5,000 smaller than the noise floor of SDR. Scaling temperature and probe signal amplitude accordingly suggests that measurements near $\approx50$~K may be possible with SDR. Measurements at lower temperatures would require more advanced and somewhat costlier SDRs. A promising candidate is SDRlab 122-16 produced by Red Pitaya \cite{pitaya}. Its sensitivity measured over the high frequency range is $-122\,\,\mathrm{dBm}\approx0.2\,\,\mu$V$_\mathrm{rms}$, which may enable measurements down to~$\approx2$~K.

\section{Conclusion and outlook}

Our work brings together the physics of nanomechanical systems and the technology of amateur radio telecommunications. We demonstrate an original approach to perform nontrivial nanomechanical measurements using software-based instrumentation that is accessible to everyone. We show that SDR dongles combined with simple data processing on a computer can be used to measure the driven response of a nanomechanical resonator based on few-layer graphene. Our measurements enable quantitative analysis of the mode dynamics. Two vibrational modes can be measured in parallel. Information encoded in modulated vibrations and transduced into light intensity modulations can be readily extracted. Looking ahead, a more elaborate and improved setup may be built upon our simple system. The computer used in our setup, an aging and sluggish laptop, may be replaced with a single board computer such as a Raspberry Pi. Rapid progress in SDR technology is expected to lower the measurement noise floor \cite{pitaya}, a prerequisite for the sensitive detection of force and mass and for all resonant measurements based on nanomechanics at cryogenic temperatures. In addition, future SDRs are expected to perform digital-to-analog conversion directly at the RF front-end and downconvert the sampled signal to baseband without the need for an IF stage, making measurements faster and cleaner. With regard to multiple channel measurements, arrays of SDRs with synchronized clocks will enable new applications such as phase-coherent multichannel transceivers \cite{Laakso2020}, which may also prove useful to study arrays of coupled nanomechanical resonators \cite{Buks2002,Sage2018,Eichler2021}.

\section*{Acknowledgments}
Joel Moser is grateful to S\'ebastien Hentz, Antoine Reserbat-Plantey, Fabien Vialla, Arnaud Ralko, Yue Ying, Zhuo-Zhi Zhang, and Xiang-Xiang Song for enlightening discussions. Joel Moser acknowledges conversations with Ike Tubin, Stephen Koss, and the Suzhou History Group about ``TaiHu Mei''. This work was supported in part by the National Natural Science Foundation of China (grant numbers 62150710547 and 62074107), in part by the International Cooperation and Exchange of the National Natural Science Foundation of China NSFC-STINT (grant number 61811530020), and in part by the project of the Priority Academic Program Development (PAPD) of Jiangsu Higher Education Institutions.

\appendix

\renewcommand{\thefigure}{A\arabic{figure}}
\setcounter{figure}{0}

\section{Frequency stability of low phase noise signal generator}

\begin{figure*}[t]
\centering
\includegraphics{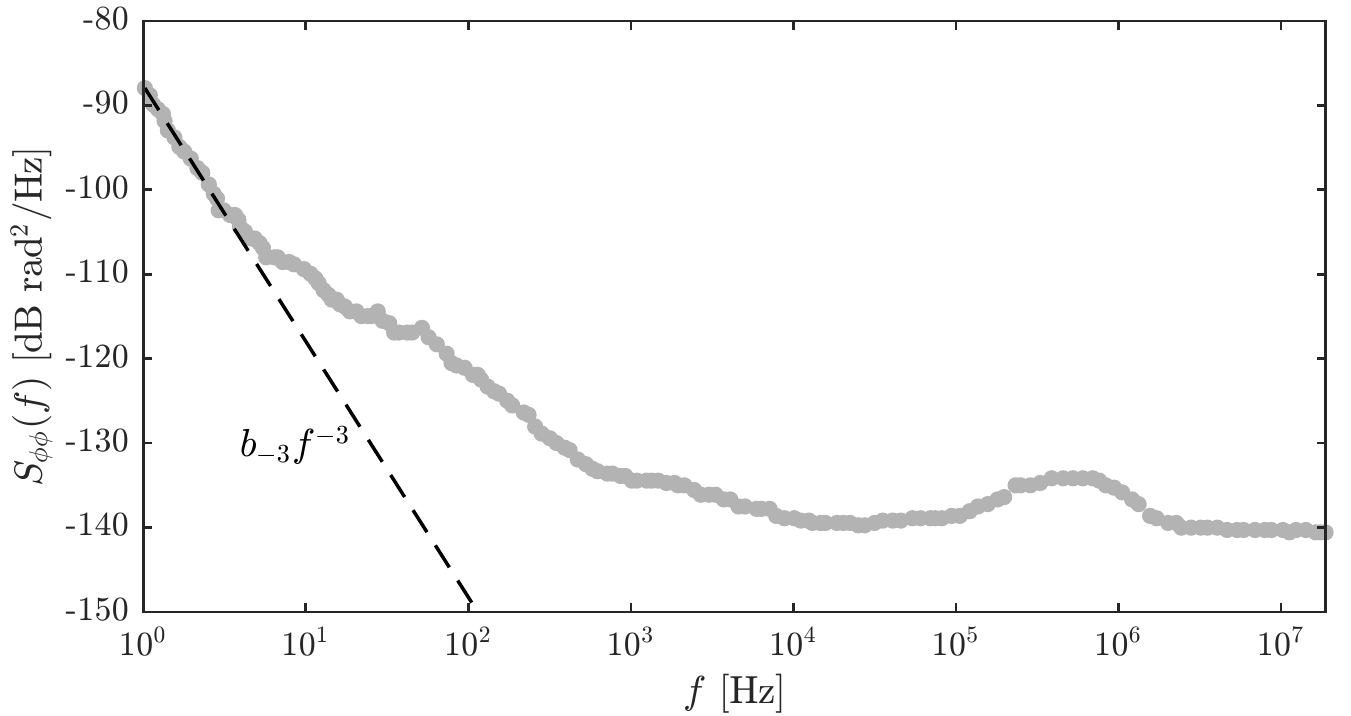}
\caption{Frequency stability of the signal generator estimated from its phase noise spectrum $S_{\phi\phi}(f)$. Dashed line shows $b_{-3}f^{-3}$.}\label{PhaseNoise}
\end{figure*}

We estimate the frequency stability of the signal generator (N5181B with low phase noise option by Keysight) from its phase noise spectrum $S_{\phi\phi}(f)$ at carrier frequency $\nu_0=100$~MHz, where $f$ is Fourier frequency (Fig.~\ref{PhaseNoise}). The frequency stability is given by the minimum value of the Allan variance of fractional frequency fluctuations, $[\sigma_y^2]_\mathrm{min}$. The latter is reached where frequency flicker dominates frequency noise \cite{Rubiola}. Frequency flicker noise and phase noise are related as $S_{\phi\phi}(f)\sim b_{-3}f^{-3}$, where $b_{-3}$ is a coefficient. We find $b_{-3}=-91\,\,\mathrm{dBc/Hz}=-88\,\,\mathrm{dB}\,\,\mathrm{rad}^2\mathrm{/Hz}$. As a result, the estimate of the frequency stability reads [ibid., Chapter~6]:
\begin{equation}
\begin{aligned}
&[\widehat{\sigma_y^2}]_\mathrm{min}=2\mathrm{ln}2\,\frac{b_{-3}}{\nu_0^2}=1.39\times\frac{10^{-88/10}}{(100\times10^6)^2}\\
&\Rightarrow\left([\sigma_y^2]_\mathrm{min}\right)^{1/2}\approx5\times10^{-13}\,,
\end{aligned}
\end{equation}
where $b_{-3}$ is in units of dB~rad$^2$/Hz, and $\widehat{\cdot}$ denotes an experimental estimate.

\section{Quantifying the dynamics of the first vibrational mode}
\begin{figure}[h]
\centering
\includegraphics[scale=0.85]{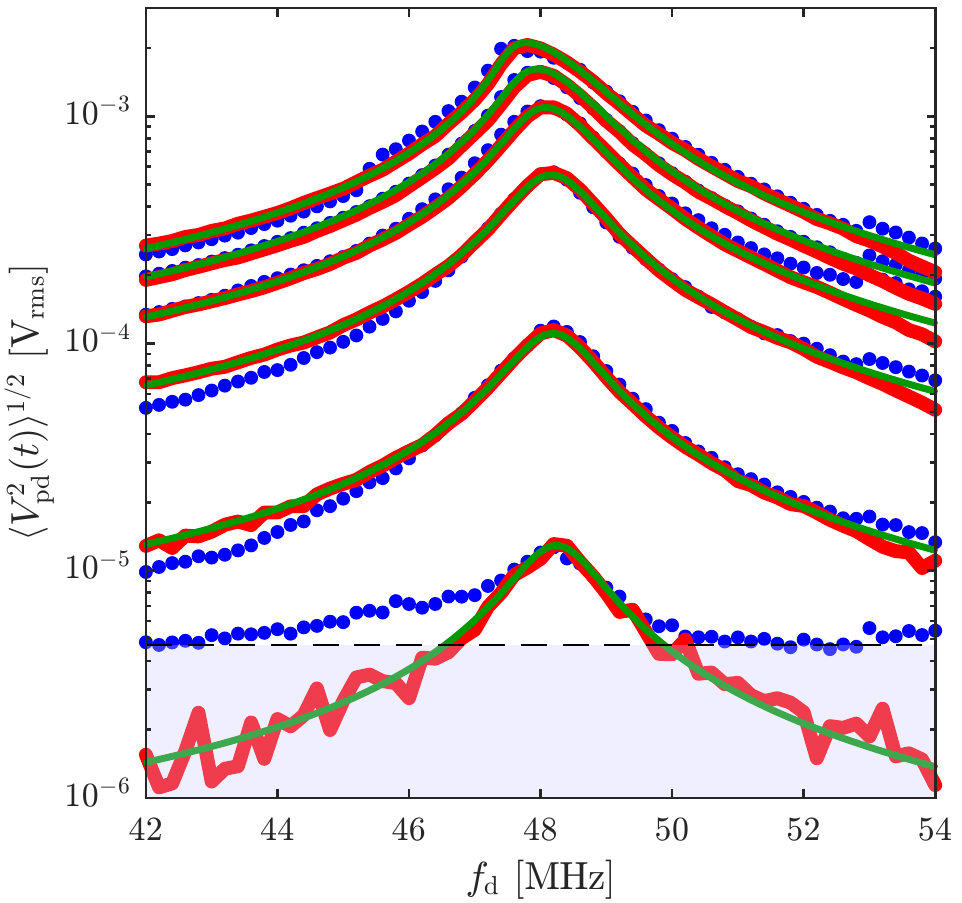}
\caption{Fitting the measured response of the first mode to Eq.~(\ref{eqn}). Peak amplitudes of driving voltage, from bottom to top: $V=0.22$, $2.24$, $11.19$, $22.39$, $33.58$ and $44.77$~mV. Gate voltage $V_\mathrm{dc}=11$~V. Horizontal dashed line and gray shading indicate the measurement background of SDR.}\label{fits}
\end{figure}
We quantify the dynamics of the first vibrational mode of our resonator by fitting its measured frequency response to Eq.~\ref{eqn}. The peak amplitude of the driving force is $F=C^\prime V_\mathrm{dc}V$, where $C^\prime\approx4\times10^{-10}$~F/m is the derivative of the capacitance between the membrane and the gate (obtained with COMSOL), $V_\mathrm{dc}$ is a dc voltage and $V$ is the peak amplitude of an oscillating voltage, both applied between the membrane and the gate. We set $V_\mathrm{dc}=11$~V. Driving voltages used in Fig.~\ref{fits}, from bottom curve to top curve, are $V=0.22$, $2.24$, $11.19$, $22.39$, $33.58$ and $44.77$~mV; these are peak voltage amplitudes which account for the full reflection of voltage waves incident on the resonator due to impedance mismatch. Blue dotted curves (resp. red curves) in Fig.~\ref{fits} show the response $\langle V_\mathrm{pd}^2(t)\rangle^{1/2}$ of the resonator as a function of $f_\mathrm{d}$ measured with SDR (resp. spectrum analyzer). Green traces show the result of fitting the response to Eq.~(\ref{eqn}). This analysis yields $k_0\approx2.1$~kg$\cdot$rad$^2/$s$^{2}$,
$\alpha\approx-(5\pm1)\times10^{15}$~kg/(m$^2\cdot$s$^2$), $\eta\approx(5\pm2)\times10^6$~kg/(m$^2\cdot$s) and $\gamma\approx(1.9\pm0.1)\times10^{-10}$~kg/s.

\begin{figure}[h]
\centering
\includegraphics[scale=0.85]{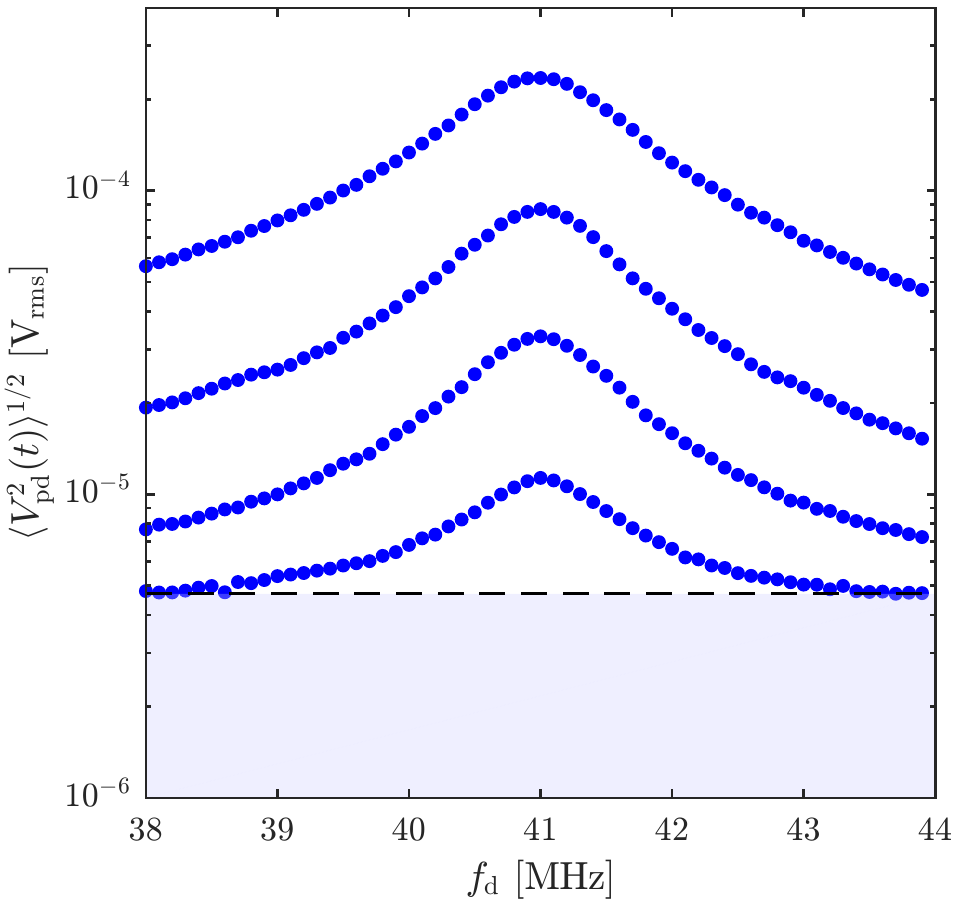}
\caption{Linear response of the first mode measured with SDR at $V_\mathrm{dc}=5$~V. Peak amplitudes of driving voltage, from bottom to top: $V=0.45$, $1.42$, $4.48$, and $14.16$~mV. Horizontal dashed line and gray shading indicate the measurement background of SDR.}\label{Vg5}
\end{figure}
The linear response of the first mode measured with SDR at $V_\mathrm{dc}=5$~V is shown in Fig.~\ref{Vg5} for various driving voltage amplitudes. Compared with the linear response at $V_\mathrm{dc}=11$~V, driving amplitudes $\approx2.2$~larger are needed here to observe resonant signals of similar strengths. This can be explained by the fact that the driving force scales as $V_\mathrm{dc}V$.

\newpage
\section{MATLAB code for parallel measurements of vibrational modes with SDR dongles}
Prior to running this code, it is necessary to open one command prompt window for each dongle and run \verb"rtl_tcp" in each window. For example, for two dongles, run
\begin{verbatim}
rtl_tcp -p 1234 -d 0
rtl_tcp -p 1235 -d 1
\end{verbatim}

\begin{lstlisting}
% Parallel measurements of two frequency channels with two SDR dongles.
% The code is based on open source drivers for RTL-SDR dongles found here:
% https://osmocom.org/projects/rtl-sdr/wiki/Rtl-sdr
% Communication with dongles through TCP is based on the code by Xianjun Jiao:
% https://www.mathworks.com/matlabcentral/fileexchange/45024-rtl-sdr-multi-dongles-based-flexible-spectrum-scanner
Nstar = 1; % start index for raw data
nq = 1; % start index for filtering
num_dongle = 2; % number of dongles
measurement_time = 2; % s
gain = [16.6 16.6]; % gain=0 means Automatic Gain Control
sample_rate = 2.048e6; % Hz; sampling rate of dongles
num_samples = measurement_time*sample_rate; % number of samples per frequency measurement and per dongle

%% Set frequency range for each dongle
freq1 = 48e6; % Hz, resonant frequency of mode 1
freq2 = 80e6; % Hz, resonant frequency of mode 2
freq_window_width = 14e6; % Hz
freq_step_size = 200e3; % Hz
freq_window = -ceil(freq_window_width/2):freq_step_size:ceil(freq_window_width/2);
freq_range1 = freq1 + freq_window;
freq_range2 = freq2 + freq_window;
freq_range = cat(1,freq_range1,freq_range2);
freq_range = freq_range';

%% Initialize two-channel signal generator
obj1 = instrfind('Type', 'visa-tcpip', 'RsrcName', 'TCPIP0::A-33600-00000.local::inst0::INSTR', 'Tag', '');
if isempty(obj1)
    obj1 = visa('NI', 'TCPIP0::A-33600-00000.local::inst0::INSTR');
else
    fclose(obj1);
    obj1 = obj1(1);
end
fopen(obj1);

%% Build low-pass filters to process output of dongles
fc1 = 100;% Hz, cut off frequency, dongle#1 output
fc2 = 100;% Hz, cut off frequency, dongle#2 output
fn = sample_rate/2; % Nyquist frequency
order = 6; % filter order
[z1,p1,k1] = butter(order,fc1/fn,'low');
[sos,g] = zp2sos(z1,p1,k1);
[z2,p2,k2] = butter(order,fc2/fn,'low');
[sos2,g2] = zp2sos(z2,p2,k2);

%% Access dongles through TCP (Xianjun Jiao)
real_count = zeros(1, num_dongle);
if ~isempty(who('tcp_obj'))
    for i=1:length(tcp_obj)
        fclose(tcp_obj{i});
        delete(tcp_obj{i});
    end
    clear tcp_obj;
end
tcp_obj = cell(1, num_dongle);
for i=1:num_dongle
    tcp_obj{i} = tcpip('127.0.0.1', 1233+i); % for dongle#i
end
for i=1:num_dongle
    set(tcp_obj{i}, 'InputBufferSize', 8*2*num_samples);
    set(tcp_obj{i}, 'Timeout', 60);
end
for i=1:num_dongle
    fopen(tcp_obj{i});
end

%% Set gain, sampling rate and center frequency for each dongle
for i=1:num_dongle
    set_gain_tcp(tcp_obj{i}, gain(i)*10);
end
for i=1:num_dongle
    set_rate_tcp(tcp_obj{i}, sample_rate);
end
for i=1:num_dongle
    set_freq_tcp(tcp_obj{i}, freq_range(1,i));
end

%% Data acquisition loop
for m=1:size(freq_window,2)
    % Set frequency of each signal generator channel
    fprintf(obj1, sprintf(strcat([';:SOUR1:FREQ ',num2str(freq_range(m,1)),';'])));
    fprintf(obj1, sprintf(strcat([';:SOUR2:FREQ ',num2str(freq_range(m,2)),';'])));
    pause(2);
    % Set center frequency of each dongle
    for i=1:num_dongle
    set_freq_tcp(tcp_obj{i}, freq_range(m,i));
    end
    pause(2);
   % Flush dongles
    for i=1:num_dongle
    fread(tcp_obj{i}, 8*2*num_samples, 'uint8');
    end
    s_all = uint8(zeros(2*num_samples,2));
   % Read out dongles (Xianjun Jiao)
    while 1
       for i=1:num_dongle
          [s_all(:, i), real_count(i)] = fread(tcp_obj{i}, 2*num_samples, 'uint8');
       end
       if sum(real_count-(2*num_samples)) = 0
          break;
       end
    end
% Convert unsigned 8-bit integer data into complex data
r = raw2iq(double(s_all));
y_real = real(r(Nstar:end,:));
y_imag = imag(r(Nstar:end,:));
y = y_real + 1i*y_imag;
t_array = linspace(0,size(y,1)-1,size(y,1))/sample_rate;

%% Filter raw I and raw Q
% Filter dongle#1 output
i_filt1 = filtfilt(sos,g,real(y(:,1)));
i_filt_segment1 = i_filt1(nq:size(i_filt1,1));
q_filt1 = filtfilt(sos,g,imag(y(:,1)));
q_filt_segment1 = q_filt1(nq:size(q_filt1,1));
y_segment_filt1 = i_filt_segment1 + 1i*q_filt_segment1;

% Filter dongle#2 output
i_filt2 = filtfilt(sos2,g2,real(y(:,2)));
i_filt_segment2 = i_filt2(nq:size(i_filt2,1));
q_filt2 = filtfilt(sos2,g2,imag(y(:,2)));
q_filt_segment2 = q_filt2(nq:size(q_filt2,1));
y_segment_filt2 = i_filt_segment2 + 1i*q_filt_segment2;

% Build time axes
t_segment_filt1 = linspace(0,size(i_filt_segment1,1)-1,size(i_filt_segment1,1))/sample_rate;
t_segment_filt2 = linspace(0,size(i_filt_segment2,1)-1,size(i_filt_segment2,1))/sample_rate;

% Compute response amplitude for the two dongles
% \bar{R1} = mean(R1), \bar{R2} = mean(R2)
R1 = abs(y_segment_filt1); % dongle#1
R2 = abs(y_segment_filt2); % dongle#2

%% plot data at current dongle frequencies
figure;
    subplot(6,1,1) % raw I(t) trace for dongle#1
    plot(t_array,real(y(:,1)),'b');
    subplot(6,1,2) % raw I(t) trace for dongle#2
    plot(t_array,real(y(:,2)),'r');
    subplot(6,1,3) % filtered I(t) trace for dongle#1
    plot(t_segment_filt1,i_filt_segment1,'b','linewidth',2);
    subplot(6,1,4) % filtered I(t) trace for dongle#2
    plot(t_segment_filt2,i_filt_segment2,'r','linewidth',2);
    subplot(6,1,5) % magnitudes of filtered (I(t), Q(t)) vectors, dongle#1
    plot(t_segment_filt1,R1,'b','linewidth',2);
    subplot(6,1,6) % magnitudes of filtered (I(t), Q(t)) vectors, dongle#2
    plot(t_segment_filt2,R2,'r','linewidth',2);
end

% close TCP
for i=1:num_dongle
    fclose(tcp_obj{i});
end
for i=1:num_dongle
    delete(tcp_obj{i});
end
clear tcp_obj;

%% Extra functions (Xianjun Jiao)
function b = raw2iq(a)
c = a(1:2:end,:) + 1i.*a(2:2:end,:);
b = c - kron(ones(size(c,1),1),(sum(c,1)./size(c,1)));
%
function tcp_obj = set_freq_tcp(tcp_obj, freq)
fwrite(tcp_obj, 1, 'uint8');
fwrite(tcp_obj, uint32(freq), 'uint32');
%
function tcp_obj = set_gain_tcp(tcp_obj, gain)
if gain
    fwrite(tcp_obj, 3, 'uint8');
    fwrite(tcp_obj, uint32(1), 'uint32');
    fwrite(tcp_obj, 4, 'uint8');
    fwrite(tcp_obj, uint32(gain), 'uint32');
else
    fwrite(tcp_obj, 3, 'uint8');
    fwrite(tcp_obj, uint32(0), 'uint32');
end
%
function tcp_obj = set_rate_tcp(tcp_obj, rate)
fwrite(tcp_obj, 2, 'uint8');
fwrite(tcp_obj, uint32(rate), 'uint32');
\end{lstlisting}

\section{Delay between two parallel measurements}

\begin{figure*}[!t]
\centering
\includegraphics{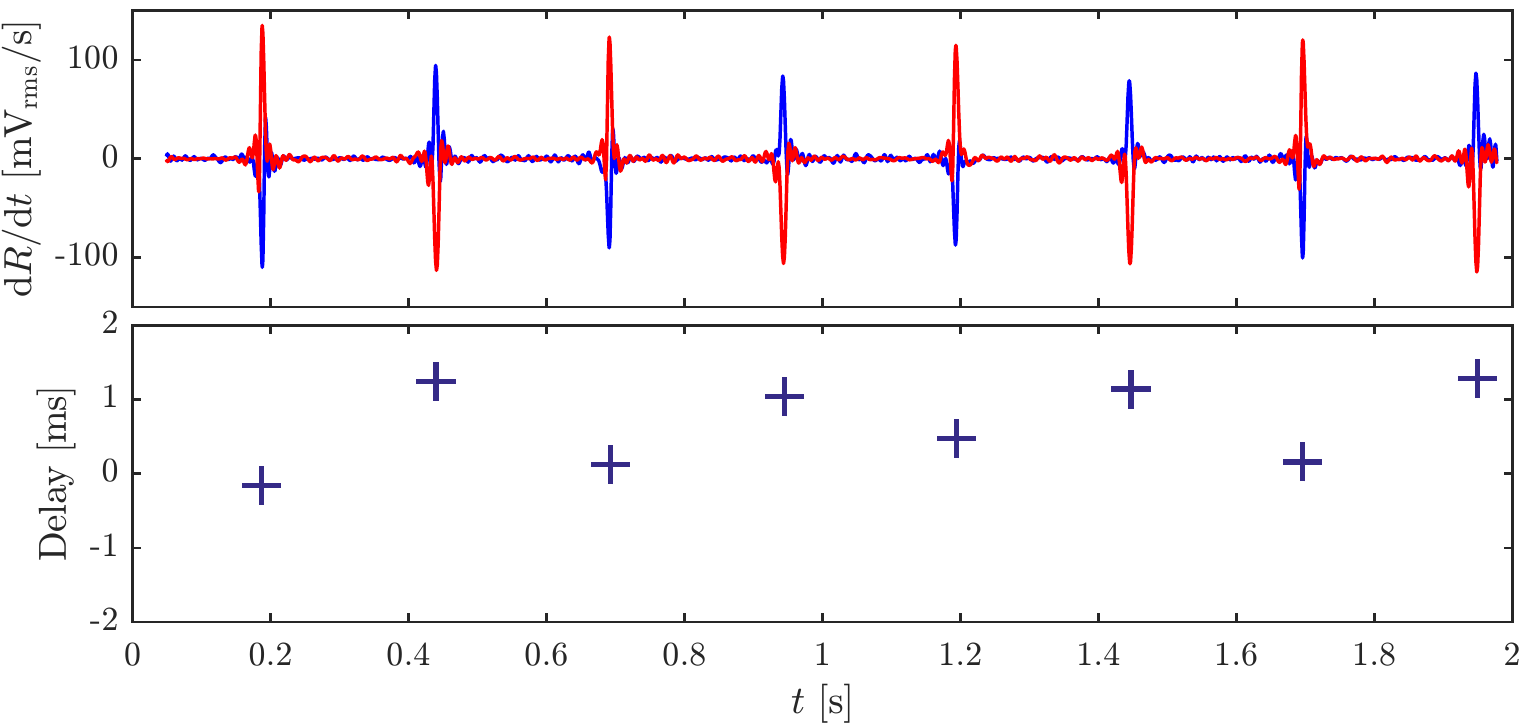}
\caption{Delay between the two parallel measurements shown in Fig.~4d of the main text. Upper panel: $\mathrm{d}R/\mathrm{d}t$ for the data in Fig.~4d, upper panel (blue traces) and for the data in Fig.~4d, lower panel (red traces). Lower panel: delay between dips in $\mathrm{d}R/\mathrm{d}t$ of one data set and peaks in $\mathrm{d}R/\mathrm{d}t$ of the other data set.}\label{delay}
\end{figure*}

To estimate the delay between two measurements made in parallel, we use the two data sets shown in Fig.~4d of the main text and plot their derivatives with respect to time $\mathrm{d}R/\mathrm{d}t$ in Fig.~\ref{delay}, upper panel. The lower panel shows the delay between a peak in $\mathrm{d}R/\mathrm{d}t$ of one data set and the nearest dip in $\mathrm{d}R/\mathrm{d}t$ of the other data set, yielding a delay of about 1~ms.

\section{Dispersion of the first mode and of the second mode}

\begin{figure}[h]
\centering
\includegraphics{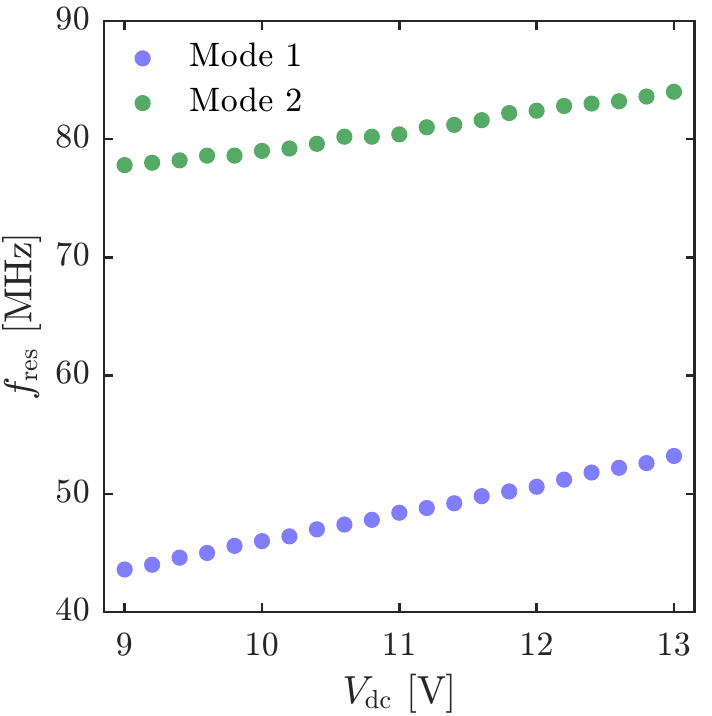}
\caption{Resonant frequency $f_\mathrm{res}=f_0$ (resp. $f_1$) of Mode~1 (resp. Mode~2) as a function of $V_\mathrm{dc}$.}\label{dispersion}
\end{figure}

The resonant frequency of Mode~1, $f_0$, and that of Mode~2, $f_1$, are shown as a function of gate voltage $V_\mathrm{dc}$ in Fig.~\ref{dispersion}. Near $V_\mathrm{dc}=11$~V, we find $\mathrm{d}f_0/\mathrm{d}V_\mathrm{dc}\approx2.4$~MHz/V and $\mathrm{d}f_1/\mathrm{d}V_\mathrm{dc}\approx1.6$~MHz/V.

\section{GNU Radio flowgraph used to demodulate frequency-modulated photodetector output}

\begin{figure*}[t]
\centering
\includegraphics{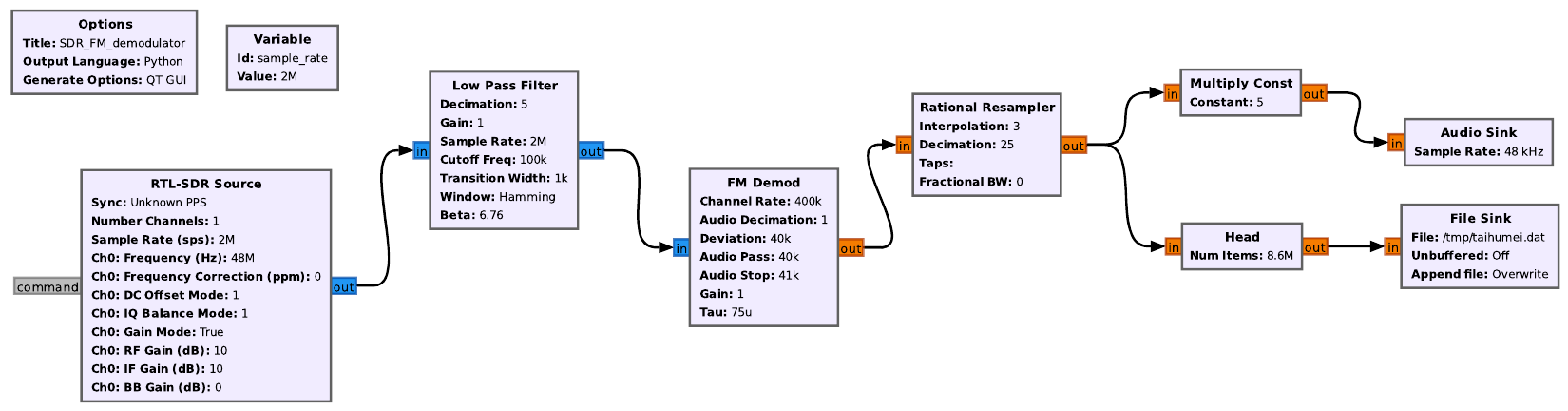}
\caption{GNU Radio flowgraph used to demodulate frequency-modulated optomechanical signals at the output of our photodetector.}\label{gnuflowgraph}
\end{figure*}

Demodulation of frequency-modulated (FM) signals is a basic function in the GNU Radio environment \cite{GNUR}. We built the flowgraph depicted in Fig.~\ref{gnuflowgraph} by following GNU Radio tutorials \cite{GNUtutorials}. We use this flowgraph to demodulate FM optomechanical signals at the output of our photodetector. Its principle of operation goes as follows. \verb"RTL-SDR Source" sets parameters for the SDR dongle, including center frequency $f_\mathrm{tuner}+f_\mathrm{IF}=48$~MHz, sampling rate $f_\mathrm{S}=2$~MHz, RF gain (before downconversion to IF), and IF gain (BB gain is for a device called Hack RF which we do not use). $I$ and $Q$ waveforms at the output of \verb"RTL-SDR Source" are low-pass filtered using \verb"Low Pass Filter" (100~kHz cutoff frequency, transition width of 1~kHz from pass-band to stop-band). A Hamming window is used to build this low-pass filter; the shape factor beta is used only with a Kaiser window. The filtered waveforms are decimated (down-sampled) by a factor of 5, so they are sampled at a rate of $2\times10^6/5=4\times10^5$~Hz. \verb"FM Demod" demodulates the filtered $I$ and $Q$ waveforms. The sampling rate of this block matches the decimated sampling rate of the low-pass filtered waveforms. We use a frequency deviation of 40~kHz that encompasses the frequency range of the analog audio signal. The demodulated waveform is low-pass filtered. A de-emphasis filter with a response time $\tau=75$~$\mu$s is used to attenuate high frequency components, so the signal-to-noise ratio of the demodulated waveform is more uniform across the audio frequency range. The demodulated waveform is decimated by \verb"Rational Resampler". The sampling rate $f_\mathrm{S}^{\mathrm{(RR)}}$ of the waveform at the output of this block is the sampling rate at the output of \verb"FM Demod" multiplied by \verb"Interpolation" and divided by \verb"Decimation", so that $f_\mathrm{S}^{\mathrm{(RR)}}=48\times10^3$~Hz. \verb"Multiply Const" scales the amplitude of the waveform at the output of \verb"Rational Resampler". \verb"Num Items" in \verb"Head" is set to the time length of the song (the analog input whose duration is about 3~min) multiplied by $f_\mathrm{S}^{\mathrm{(RR)}}$. \verb"Audio Sink" sends the resulting waveform to the sound card of the computer at a rate of 48~kHz~$=f_\mathrm{S}^{\mathrm{(RR)}}$. \verb"File Sink" saves the audio file in binary format. We convert the latter into wav format with MATLAB, then we convert the .wav file into mp3 format using Audacity \cite{Audacity}.

\section{Audio files for demodulated optomechanical signals}

Audio files in mp3 format accompany this work.
\begin{itemize}
\item{\href{https://drive.google.com/file/d/195MeBk_nUGqFRhv0lnGM4luWAXcNxY1N/view?usp=sharing}{TaihumeiOriginal.mp3} is the original recording of Ms. Jue Wang's performance of `Taihu Mei'.}
\item{\href{https://drive.google.com/file/d/1I8mLFCAv-iTB0pTrM90mIv6HTPql9-LF/view?usp=sharing}{TaihumeiDemodulated.mp3} is the demodulated `Taihu Mei' optomechanical signal.}
\item{\href{https://drive.google.com/file/d/1KKZfFiUZOq5yr791wfz4s0q9eWk7FD8n/view?usp=sharing}{ShakespeareDemodulated.mp3} is the demodulated optomechanical signal based on poetry. The message in the modulated input signal was William Shakespeare' Sonnet~116. It was recited by Sir Laurence Olivier on The Dick Cavett Show in 1973. The original recording was posted on YouTube by its owners and is available here: \url{https://youtu.be/kWDCCf1CYXI}.}
\item{\href{https://drive.google.com/file/d/1yKaf1xWOaTZ3-tNzk9cgXa8X8BVrwTnL/view?usp=sharing}{MusicalDemodulated.mp3} is the demodulated optomechanical signal based on a musical. The message in the modulated input signal was an excerpt from the song ``Singin' in the Rain'', performed by Gene Kelly, with lyrics by Arthur Freed and music by Nacio Herb Brown. The original recording was posted on YouTube by its owners and is available here: \url{https://youtu.be/swloMVFALXw}.}
\end{itemize}

\end{document}